\documentclass[12pt]{article}
\usepackage[export]{adjustbox}
\usepackage{graphicx}
\usepackage{graphics}
\usepackage{epsfig}
\usepackage{amssymb}
\usepackage{amsmath}
\usepackage{color}
\usepackage{textcomp}
\usepackage{latexsym}
\usepackage{physics}

\usepackage[margin=2.5cm,footskip=0.6cm]{geometry}
\begin{document}

\begin{center}
{\Large{\bf Derivation and Analysis of Amplitude Equation for Generalized AMB+ in Presence of Chemical Reaction}} \\
\ \\
by \\
Sayantan Mondal and Prasenjit Das\footnote{prasenjit.das@iisermohali.ac.in} \\
Department of Physical Sciences, Indian Institute of Science Education and Research -- Mohali, Knowledge City, Sector 81, SAS Nagar 140306, Punjab, India. \\
\end{center}

\begin{abstract}
\noindent We derive and analyze the amplitude equation for the roll patterns in case of generalized Active Model B+ (AMB+) in the presence of chemical reactions. The generalized AMB+ differs from the original AMB+ introduced by Tjhung \textit{et al.} [E. Tjhung \textit{et al.}, Phys. Rev. X \textbf{8}, 031080 (2018)] by the addition of a quadratic term, $g\phi^2$, in the expression for the equilibrium part of the current. Also, the model includes a rotation-free active current of strength $\lambda$ and a rotational current of strength $\xi$. The inclusion of a chemical reaction with rate $\Gamma$ removes the conservation constraint and introduces a preferred wavenumber that governs the pattern formation below a critical reaction rate $\Gamma_c$. We argue for the analytical form of the amplitude equation based on symmetry considerations and explicitly derived it using multiscale analysis. By taking different limits of $g$, $\lambda$, and $\xi$, we recover amplitude equations for several well-known physical models as special cases and determine the nature transitions close to the onset of instability. We find that for $g = 0$, the transition is always supercritical, whereas for $g \ne 0$, the transition between the supercritical and subcritical regimes depends sensitively on the model parameters. Further, we derive the condition for the \textit{Eckhaus instability} from the stability analysis of the amplitude equation as well as from the phase diffusion equation, and find that it is independent of $g$.
\end{abstract}

\newpage
\section{Introduction}\label{sec1}
Pattern formation in active-matter systems has received enormous attention in recent years~\cite{S10,MJSTJR13,JGMJC22}. In these systems, individual units exhibit persistent motion arising from the conversion of internal energy or from energy fluxes in their surroundings into directed motion—a property known as motility. This motility causes violation of detailed balance at the microscopic level and enables active systems to develop self-organized spatiotemporal patterns across multiple length scales. Examples range from bacterial colonies (with constituent sizes of a few $\mu$m) to schools of fish and flocks of birds (a few cm), and further to herds of sheep and coordinated movements in human crowds (a few m)~\cite{DMA18,AI24,MARYHW23,AF23}.

Over the past few decades, pattern formation in active systems has been investigated experimentally, through particle-based simulations, and theoretically using coarse-grained models. One of the earliest experimental study on self-assembly in bacterial systems was due to Dombrowski \textit{et al}~\cite{CLSRJ04}. They observed large-scale coherent structure formation in bacterial dynamics within a confined geometry. Subsequently, experimental studies on synthetic active particles, such as Janus particles, demonstrated that various self-assembled structures can be achieved by tuning the surrounding environment and confinement. Deseigne \textit{et al.} observed large-scale collective motion in a monolayer of vibrated disks with built-in polar asymmetry~\cite{J0H10}. These driven asymmetric disks effectively mimic self-propelled particles. Later, Buttinoni \textit{et al.} reported dynamical clustering and phase separation in suspensions of self-propelled colloidal particles~\cite{IJFHCT13}. B\"auerle \textit{et al.} studied the self-organization of light-activated active particles whose motility depends on the density of their surroundings~\cite{TATC18}.

The first particle-based simulation model for active matter was proposed by Vicsek \textit{et al}~\cite{TAEIO95}. Their study on self-propelled particles shows the emergence of collective motion purely due to the motility, even in the absence of attractive interactions. This model, known as the Vicsek model, was later extended to incorporate various forms of mutual interactions. For example, Redner \textit{et al.} studied the structure and dynamics of an active colloidal system in which particles interact via a WCA potential in the radial direction, while their angular dynamics are governed by random noise~\cite{GMA13}. Later, many realistic molecular simulations were performed that incorporated mutual interactions in both the radial and angular directions. For example, alignment interactions for symmetric particles, such as spheres, can be modeled using Kuramoto-type couplings~\cite{AV2010}, whereas for asymmetric particles, such as rods, simple collisions suffice to generate angular interactions~\cite{DDH2010}. These models have also been further extended to account for chiral motion of the particles~\cite{LPL2019}. In all these models, depending on the model parameters, the systems exhibit rich variety of static and dynamic patterns, as well as transitions between them. For example, Zhang and Fodor observed a wide variety of dynamical patterns -- ranging from spiral waves to defect turbulence -- in the context of pulsating active matter~\cite{YE23}.

In recent years, pattern formation in active matter systems has also been extensively studied using coarse-grained models~\cite{JM13,STMAC2013,TJAH14,SP24,LJSU20}. Some of these models have been proposed phenomenologically, while others have been derived directly by coarse-graining the microscopic dynamics described above. These models play a central role in studies of pattern formation, serving as a bridge between microscopic details and continuum descriptions of the system in terms of the dynamics of slowly varying fields (e.g., density). In this context, Active Model B~(AMB)~\cite{RAJRDM14,SSS21} and its extension, Active Model B+~(AMB+)~\cite{ECM18,PSS25}, serve as minimal yet versatile models for scalar active systems. AMB generalizes the passive Model B~(MB) by incorporating activity-induced modifications to the chemical potential that break time-reversal symmetry~(TRS). In AMB+, a non-gradient term is added to the current of AMB, leading to persistent rotational currents and enabling richer dynamical phenomena such as arrested phase separation and microphase formation. Apart from these models, Tailleur and Cates derived a dynamical equation for the density field by coarse-graining the run-and-tumble dynamics, enabling the study of domain formation through ‘self-trapping’ and other collective phenomena~\cite{JM08}. Subsequently, several groups extended this framework to coarse-grain different microscopic models and to explore the underlying mechanisms of pattern formation~\cite{ANYCAJJ20,JJ20,AJAYPJ14}. Notably, the works of Saha \emph{et~al.} and You \emph{et~al.} on nonreciprocal active matter demonstrate the emergence of traveling density waves~\cite{SJR20,ZAM20}.

In pattern formation, the focus lies on the mechanisms of pattern selection and their subsequent evolution. In this context, the analysis of the amplitude equation provides a powerful framework~\cite{MP93,R06}. It enables us to capture a range of generic features of emerging patterns across diverse physical systems, particularly near the onset of transition, where a uniform base state undergoes a finite-wavelength instability. Far from the instability, however, the pattern dynamics and properties depend sensitively on the specific details of the system. For example, the reaction–diffusion and Swift–Hohenberg~(SH) equations possess very different mathematical structures, yet the amplitude equations governing their stationary stripe solutions are identical in form~\cite{A52,BCADA03,JE06}. More precisely, the form of the amplitude equation depends on the nature of the instability at the transition point. For stationary instabilities, the amplitude obeys the real Ginzburg–Landau (GL) equation, whereas for oscillatory instabilities, it follows the complex GL equation~\cite{DP04,L06,PW09}.

The amplitude equation framework has been applied to study pattern formation in a wide range of passive systems — from Hopf and Turing instabilities in reaction–diffusion systems to hydrodynamic instabilities in fluids and other forms of spatiotemporal pattern formation~\cite{M80,ATJ93,Y97,BC00,BC01,TT03}. The dynamics of both nonconserved and conserved variables have been considered within this framework~\cite{PG90,PS20,SP03,RJF20}. Multiscale analysis has been widely used to derive amplitude equations from the underlying dynamical equations~\cite{MH09}. In contrast, similar studies for active systems are relatively scarce. For instance, Cates \emph{et~al.} derived the amplitude of the mode at the onset of MIPS in the presence of a logistic birth–death process~\cite{JCAMDAW12}. Subsequently, Zimmermann \emph{et~al.} showed that phase separation in active systems with a conservation constraint can be described by the Cahn–Hilliard equation near the onset of instability by deriving the corresponding amplitude equation~\cite{FLW18}. Frohoff-H\"ulsmann \emph{et~al.} obtained amplitude equation for a non-reciprocally coupled two-field Cahn–Hilliard system~\cite{TUL23,GT24}. Frohoff-H\"ulsmann and Thiele derived an amplitude equation describing pattern formation near a conserved-Hopf bifurcation—a large-scale oscillatory instability in systems with conservation laws~\cite{FU23}. They obtained nonreciprocal two-component Cahn–Hilliard model as a universal amplitude equation that captures a broad class of oscillatory, traveling, and localized states which are relevant in active and reactive systems.

In this paper, we systematically derive the amplitude equation for the generalized AMB+ in the presence of chemical reaction using the multiscale analysis technique. The generalized AMB+ differs from the ordinary AMB+ by the addition of a quadratic term ($g\phi^2$) of order parameter $\phi$ in the expression of chemical potential. This term arises in the case of off-critical quenches of binary mixtures.~\cite{amy85}. It appears in the modeling of liquid crystals~\cite{PJ93,P98}. In the context of active phase separation, such a term has been included in recent studies by Speck \textit {et al.}~\cite{TJAH14} and others~\cite{LFW19,T22}. The incorporation of chemical reactions introduces a preferred wavenumber that governs the pattern formation and removes the conservation constraint~\cite{B76,SDN94,SEM95,SH94}. Our goal is to explore various limits of the amplitude equation by varying $g$ and other model parameters, and to analyze their roles in determining the nature of the transitions near the onset of instability.

Given this background, the paper is organized as follows. In Sec.~\ref{sec2}, we introduce the generalized AMB+ and discuss its linear stability. A derivation of the amplitude equation using symmetry arguments is provided in Sec.~\ref{sec3}, and using multiscale analysis in Sec.~\ref{sec4}. We analyze the stability of the amplitude equation, which yields the \textit{Eckhaus instability} condition in Sec.~\ref{sec5}. Next, we derive the phase diffusion equation in Sec.~\ref{sec6}. Finally, we summarize our findings in Sec.~\ref{sec7}. 

\section{Details of Generalized AMB+}\label{sec2}
We begin by defining the space- and time-dependent order parameter $\phi(\vec r, t)$, which characterizes phase separation in any AB binary mixture. If $\rho_i(\vec r, t)$ denotes the density of species $i\in \{A,B\}$ at position $\vec r$ and time $t$, then $\phi(\vec r, t) = \rho_{\rm A}(\vec r, t) - \rho_{\rm B}(\vec r, t)$, i.e., the local concentration difference between the two species. For AMB+, the evolution equation for $\phi(\vec r, t)$ is given by
\begin{eqnarray}
\label{eqn1}
\frac{\partial \phi}{\partial t} = -\vec\nabla\cdot\vec J(\vec r,t) = -\vec\nabla\cdot\left[ -\vec\nabla\{\mu_{\rm eq} + \lambda (\vec \nabla \phi)^2\} + \xi\nabla^2\phi \vec\nabla\phi \right],
\end{eqnarray}
where $\vec{J}(\vec{r}, t)$ is the current and $\mu_{\rm eq}(\vec{r}, t)$ is the equilibrium chemical potential, given by
\begin{eqnarray}
\label{eqn2}
\mu_{\rm eq}(\vec{r},t)=-a\phi + u\phi^3 - K\nabla^2\phi,
\end{eqnarray}
which can be derived from the following free-energy functional:
\begin{eqnarray}
\label{apeqn3}
\mathcal{F}[\phi(\vec{r},t)] = \int d\vec{r} \left\{ -\frac{a}{2} \phi^2(\vec{r},t) + \frac{u}{4} \phi^4(\vec{r},t) + K\left[\vec\nabla \phi(\vec{r},t)\right]^2\right\}.
\end{eqnarray}
Here $a$, $u$ and $K$ are positive constants. The equilibrium current $\vec{J}_0(\vec{r}, t)$ can be obtained as $\vec{J}_0(\vec{r}, t) = -\vec{\nabla} \mu_{\rm eq}$. The term $\vec J_1(\vec{r},t) = -\lambda \vec\nabla(\vec \nabla \phi)^2$ represents the rotation-free active current of strength $\lambda$, which explicitly breaks TRS. The last term, $\vec J_2(\vec{r},t) = \xi\nabla^2\phi \vec\nabla\phi$ in Eq. (\ref{eqn1}) represents the rotational current of strength $\xi$. AMB+ is one of the most recently established and widely accepted models describing phase separation in active binary mixtures.

We generalize the AMB+ by adding a quadratic term $g\phi^2$ in the expression of $\mu_{\rm eq}(\vec{r},t)$. Such a term naturally arises in off-critical quenches of passive systems. Recently, coarse-graining of active matter systems Speck \emph{et~al.} has also revealed a similar term~\cite{T22}. Thus the Eq. (\ref{eqn1}) reduces to 
\begin{eqnarray}
\label{apeqn1}
\frac{\partial \phi}{\partial t} = \nabla^2  \left( -a \phi + g \phi^2 +u \phi^3- K \nabla^2 \phi  +\lambda (\vec\nabla\phi)^2\right) -\xi \vec{\nabla}\cdot\left(\nabla^2\phi \vec\nabla\phi\right).
\end{eqnarray}
Here, the parameters $g$, $\lambda$ and $\xi$ can take both positive and negative values. Equation~(\ref{apeqn1}) corresponds to the AMB proposed by Wittkowski \emph{et~al.} when $g = 0$, $\lambda \ne 0$, and $\xi = 0$. It reduces to the asymmetric Cahn–Hilliard equation for $g \ne 0$, $\lambda = 0$, and $\xi = 0$. When $g \ne 0$, $\lambda = 0$, and $\xi \ne 0$, we refer to Eq.~(\ref{apeqn1}) as the generalized AMB.

Next, we consider a reversible chemical reaction process in which species A can convert to species B and vice versa. The conversion occurs at an equal rate $\Gamma$ in both directions, represented as
\begin{eqnarray}
\label{apeqn2}
\text{A} \overset{\Gamma}{\underset{\Gamma}{\rightleftharpoons}} \text{B}.
\end{eqnarray}
Using the framework of chemical rate equations, we modify Eq.~(\ref{apeqn1}) to incorporate the effect of the reaction as follows:
\begin{eqnarray}
\label{apeqn3a}
\frac{\partial \phi}{\partial t} = \nabla^2  \left[ -a \phi + g \phi^2 +u \phi^3- K \nabla^2 \phi  +\lambda (\vec\nabla \phi)^2\right] -\xi \vec{\nabla}\cdot\left(\nabla^2\phi \vec\nabla\phi\right) - \Gamma \phi.
\end{eqnarray}
This equation serves as the starting point for studying pattern formation in binary active matter systems and forms the basis for the multiscale analysis through which we derive the corresponding amplitude equation near the onset of instability.

\subsection{Linear Stability Analysis}\label{sec2a}
We perform a linear stability analysis of Eq.~(\ref{apeqn3}) by expanding $\phi(\vec{r}, t)$ around the base state $\phi_0 = 0$ as $\phi(\vec{r}, t) = \phi_0 + \delta\phi(\vec{r}, t)$. This yields
\begin{eqnarray}
\label{apeqn4}
\frac{\partial \delta\phi}{\partial t} = \nabla^2 \left( -a\delta\phi - K \nabla^2 \delta\phi \right) - \Gamma\delta\phi.
\end{eqnarray}
Next, we introduce the Fourier decomposition of $\delta\phi(\vec{r}, t)$ as
\begin{eqnarray}
\label{apeqn5}
\delta\phi(\vec{r}, t) = \int d\vec{q} e^{i \vec{q} \cdot \vec{r}} \delta\phi(\vec{q}, t),
\end{eqnarray}
where $\delta\phi(\vec{q}, t) \propto e^{\sigma(q)t}$. Using this form in Eq.~(\ref{apeqn4}) gives the linear dispersion relation
\begin{eqnarray}
\label{apeqn6}
\sigma(q) = a q^2 - K q^4 - \Gamma.
\end{eqnarray}

\begin{figure}
\centering
\includegraphics*[width=0.60\textwidth]{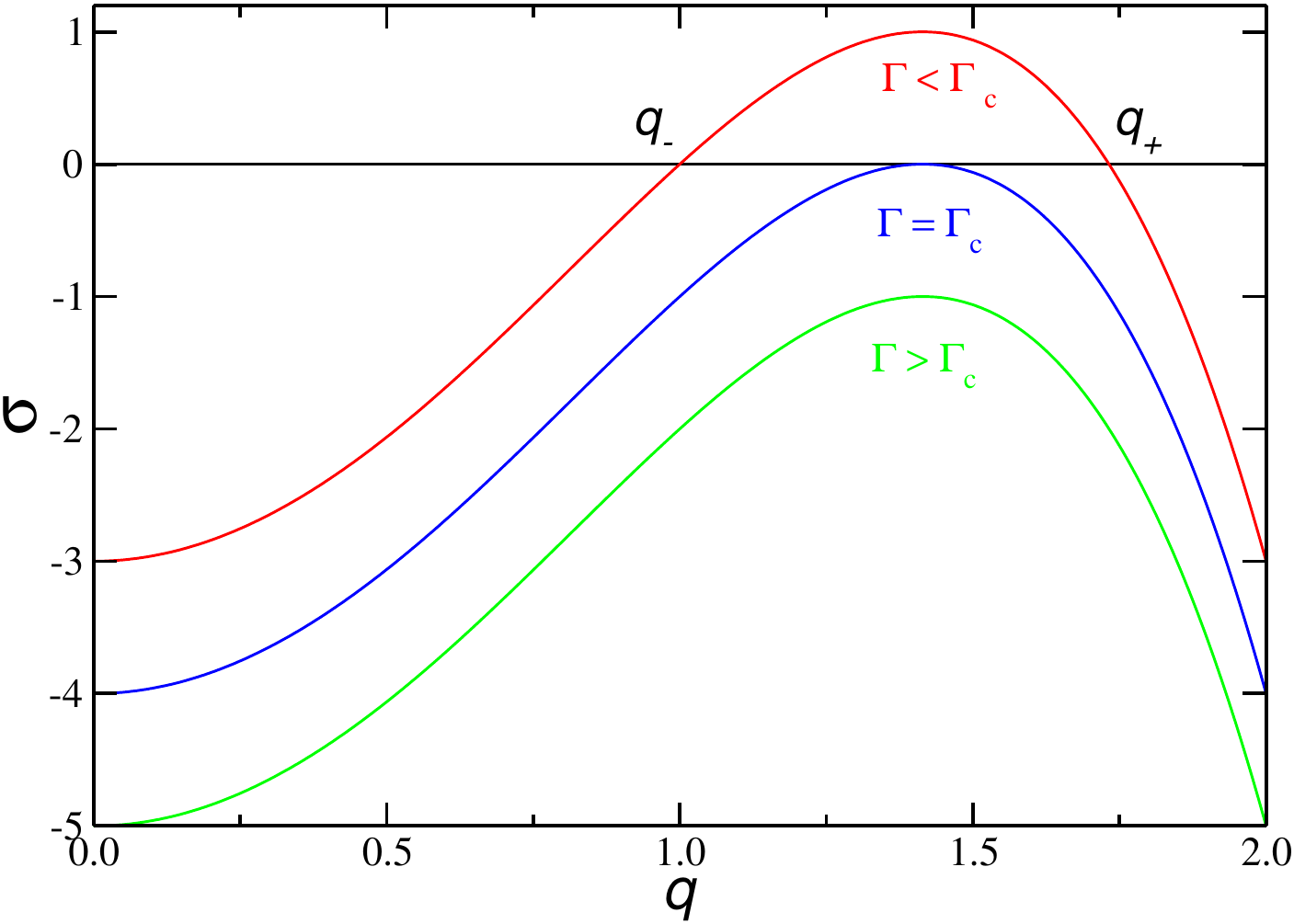}
\caption{\label{fig1} Plot of $\sigma(q)$ versus $q$, given by Eq.~(\ref{apeqn6}), for $a = 4$ and $K = 1$, at different values of $\Gamma$. Here, $\Gamma_c=4$ is the critical reaction rate below which $\sigma(q)$ becomes positive for a band of wave numbers $q\in[q_-, q_+]$. }
\end{figure}

Figure~\ref{fig1} shows the plot of $\sigma(q)$ versus $q$ for different values of $\Gamma$. The values of the other parameters are provided in the figure caption. We define $\Gamma_c = \tfrac{a^2}{4K}$ as the critical reaction rate above which $\sigma(q)$ remains negative for all $q$, indicating that any fluctuation decays in time and the homogeneous base state $\phi_0$ is stable. At $\Gamma \lesssim \Gamma_c$, the mode with wave number $q_c = \sqrt{\tfrac{a}{2K}}$ first becomes unstable, thereby setting the characteristic length scale of the emerging pattern. Furthermore, the wave number corresponding to the maximum growth rate is obtained from $\tfrac{d\sigma(q)}{d (q^2)} = 0$, which yields $q_m = \sqrt{\tfrac{a}{2K}}$, identical to the critical wave number $q_c$ at the onset of instability. For $\Gamma < \Gamma_c$, there exists a band of wave numbers between $q_-$ and $q_+$, centered around $q_c$ for which $\sigma(q) > 0$, implying that modes with $q \in [q_-, q_+]$ grow in time. The values of $q_\pm$ are obtained by setting $\sigma(q) = 0$, which gives
\begin{eqnarray}
\label{apeqn7}
q_{\pm} = \sqrt{\frac{\tfrac{a}{K} \;\pm\; \sqrt{\left(\tfrac{a}{K}\right)^{2} - \tfrac{4\Gamma}{K}}}{2}}.
\end{eqnarray}

\begin{figure}
\centering
\includegraphics*[width=0.56\textwidth]{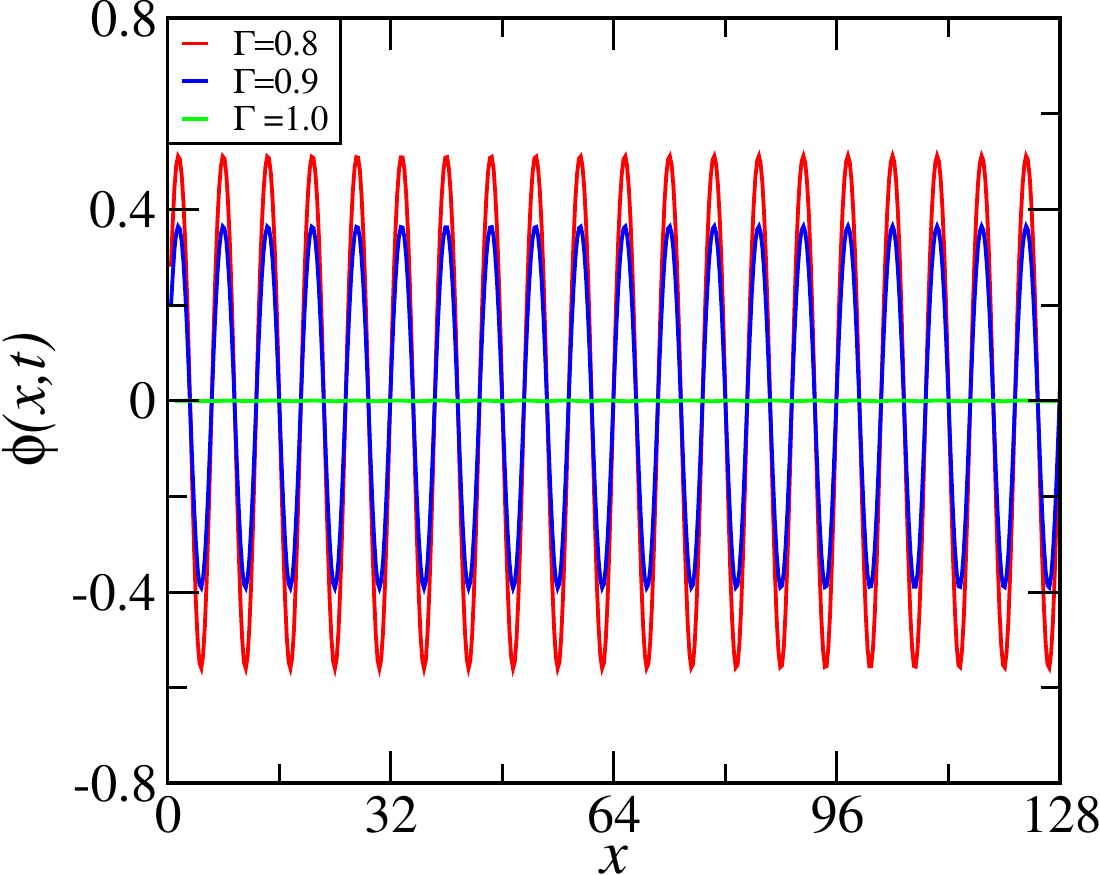}
\caption{\label{fig2} Plot of the stationary $\phi(x,t)$ versus $x$ at $t = 1000$ for a system of size $L = 128$, obtained by solving Eq.~(\ref{apeqn3a}) in $d=1$ with parameters $a = 2$, $g = 0.4$, $\lambda = 0.05$, $\xi = 0.05$, $u=1$ and $K = 1$, for different values of $\Gamma$. For this choice of parameters, $\Gamma_c = 1$ denotes the critical reaction rate.}
\end{figure}

For $\Gamma < \Gamma_c$, $q_{\pm}$ are real, and hence we expect stationary patterns in this regime. Figure~\ref{fig2} shows the stationary profile of $\phi(x,t)$ versus $x$, obtained by solving Eq.~(\ref{apeqn3a}) in $d = 1$ for different values of $\Gamma$. We use the Euler discretization method with a spatial discretization size of $\Delta x = 0.5$, and a time step of $dt = 0.0001$, which provides the required numerical stability. In all our simulations, the initial condition for the order parameter consists of random fluctuations around zero, i.e., $\phi(x,0)=\pm0.01$. We employ periodic boundary condition in the spatial direction. The fixed values of the other parameters are provided in the figure caption. Close to $\Gamma = \Gamma_c$, the amplitude of $\phi(x,t)$ is nearly zero and increases as $\Gamma$ is reduced below $\Gamma_c$.

\section{Amplitude Equation From Symmetry Arguments}\label{sec3}
The structure of the amplitude equation can be deduced directly from the underlying symmetries of the dynamical equation, without performing a full multiscale analysis. The amplitude equation inherits the same set of symmetries as the dynamical equation from which it is derived. Since the generalized AMB+ model is invariant under spatial translation and inversion symmetry operations, the corresponding amplitude equation is also expected to exhibit invariance under these symmetries.

Near the onset of instability, the general solution for $\phi(x,t)$ corresponding to a roll pattern in $d = 1$ can be approximated as 
\begin{eqnarray}
\label{apeqn8}
\phi(x,t) \sim A(X,T) e^{iq_cx} + A^*(X,T) e^{-iq_cx}.
\end{eqnarray}
where, $A(X,T)$ is the slowly varying amplitude of the harmonic $e^{iq_cx}$, evolving over much longer spatial and temporal scales $(X,T)$, as elaborated in the next section. $A^*(X,T)$ is the complex conjugate of $A(X,T)$. Under a translation $x \rightarrow x + x_0$, the amplitude $A(X,T)$ transforms as
\begin{eqnarray}
\label{apeqn9}
A(X,T) \rightarrow A(X,T) e^{iq_cx_0}.
\end{eqnarray}
Since the amplitude equation must respect translational symmetry of the original equation, its form should remain invariant under this transformation. This constraint prohibits nonlinear terms such as $A^2$, $A^3$, or $A^{*3}$ and allows only cubic nonlinearities of the form $\lvert A \rvert^2A$. Consequently, the most general form of the amplitude equation is written as
\begin{eqnarray}
\label{apeqn10}
\frac{\partial A}{\partial T}= \alpha_1 A + \alpha_2 \lvert A \rvert^2 A + \alpha_3 \frac{\partial^2 A}{\partial X^2}.
\end{eqnarray}
Here, the coefficients $\alpha_i$ ($i=1,2,3$) are, \textit{a priori} considered to be complex. 

Next, the inversion symmetry $X \rightarrow -X$ implies the transformation $A \rightarrow A^{*}$, which leaves the general solution unchanged. Since the amplitude equation~(\ref{apeqn10}) must also respect inversion symmetry, this requirement enforces $\alpha_i = \alpha^{*}_i$, i.e., all coefficients must be real. The exact values of these coefficients can be obtained through a multiscale analysis, which we present in the next section.

\section{Derivation Using Multiscale Analysis}\label{sec4}
To derive the amplitude equation, we begin by introducing a small parameter $\epsilon$ defined as $\epsilon^2 = \frac{\Gamma_c - \Gamma}{\Gamma_c}$. The parameter $\epsilon$ thus measures the distance from the onset of instability and serves as the control parameter for the multiscale analysis. Rewriting Eq.~(\ref{apeqn7}) in terms of $\epsilon$ gives
\begin{equation}
\label{apeqn11}
q_{\pm} = q_c \left[ \sqrt{1 \pm \sqrt{1 - \frac{\Gamma}{\Gamma_c}} } \right]
= q_c \left[ \sqrt{1 \pm \epsilon } \right].
\end{equation}
Hence, the width of the band of unstable wave numbers, $\Delta q = q_{+} - q_{-}$, behaves as $\Delta q \sim\epsilon$.

At the instability threshold, where  $q\simeq q_c$, the growth rate $\sigma$ becomes
\begin{eqnarray}
\label{apeqn13}
\sigma \simeq a q_c^2 + K q_c^4 - \Gamma_c(1-\epsilon^2)\sim \mathcal{O}(\epsilon^2).
\end{eqnarray}
Thus, the dynamics evolve slowly in the vicinity of the instability, and it is appropriate to introduce rescaled slow variables $X = \epsilon x$ and $T = \epsilon^2 t$ to capture the long-wavelength spatial variations and slow temporal evolution of the system near the threshold.

The differential operators connecting fast and slow scales are as follows
\begin{equation}
\label{apeqn14}
\frac{\partial}{\partial t} \rightarrow  \frac{\partial}{\partial t} + \epsilon^2 \frac{\partial}{\partial T},
\end{equation}
\begin{equation}
\label{apeqn15}
\frac{\partial}{\partial x} \;\rightarrow\; \frac{\partial}{\partial x} + \epsilon \frac{\partial}{\partial X},
\end{equation}
\begin{equation}
\label{apeqn16}
\frac{\partial^2}{\partial x^2} \;\rightarrow\; \frac{\partial^2}{\partial x^2}
+ 2 \epsilon \frac{\partial}{\partial x} \frac{\partial}{\partial X}
+ \epsilon^2 \frac{\partial^2}{\partial X^2},
\end{equation}
\begin{equation}
\label{apeqn17}
\frac{\partial^4}{\partial x^4} \;\rightarrow\; \frac{\partial^4}{\partial x^4}
+ 4 \epsilon \frac{\partial^3}{\partial x^3} \frac{\partial}{\partial X}
+ 6 \epsilon^2 \frac{\partial^2}{\partial x^2} \frac{\partial^2}{\partial X^2}+\mathcal{O}(\epsilon^3)
\end{equation}
Here, we kept terms up to order $\epsilon^2$. Further, we expand $\phi(x,t)$ in orders of $\epsilon$ as
\begin{align}
\phi(x,t)=&\epsilon \phi_1 + \epsilon^2 \phi_2 + \epsilon^3 \phi_3 +\mathcal{O}(\epsilon^4) \nonumber \ \\ 
\label{apeqn19}
=&\epsilon [A(X,T) e^{iq_c x} + A^*(X,T) e^{-iq_c x}] + \epsilon^2 [B(X,T) e^{i2q_c x} + B^*(X,T) e^{-i2q_c x}]  \nonumber \\ 
& + \epsilon^3 [C(X,T) e^{i3q_c x} + C^*(X,T) e^{-i3q_c x}] + \mathcal{O}(\epsilon^4).
\end{align}

Next, we substitute Eqs.~(\ref{apeqn14})–(\ref{apeqn19}) into Eq.~(\ref{apeqn3a}). The LHS of Eq.~(\ref{apeqn3a}) then transforms as
\begin{align}
\label{n1}
\frac{\partial\phi}{\partial t}\rightarrow \left(\frac{\partial}{\partial t} + \epsilon\frac{\partial}{\partial T}\right) \left(\epsilon \phi_1 + \epsilon^2 \phi_2 + \epsilon^3 \phi_3 +\mathcal{O}(\epsilon^4)\right).
\end{align}
Similarly, the first term on the right-hand side becomes
\begin{align}
\label{n2}
-a\frac{\partial^2\phi}{\partial x^2}\rightarrow -a\left(\frac{\partial^2}{\partial x^2}
+ 2\epsilon\frac{\partial}{\partial x}\frac{\partial}{\partial X} +\epsilon^2\frac{\partial^2}{\partial X^2} \right)\left(\epsilon \phi_1 + \epsilon^2 \phi_2 + \epsilon^3 \phi_3 +\mathcal{O}(\epsilon^4)\right).
\end{align}
We proceed in an analogous manner for the remaining terms of Eq.~(\ref{apeqn3a}). We define a linear operator $L$ as
\begin{equation}
\label{apeqn18} 
L=\left(-a\frac{\partial^{2}}{\partial x^{2}}-K\frac{\partial^{4}}{\partial x^{4}}-\Gamma_c\right),
\end{equation}
which will reappear in successive orders of $\epsilon$ in our multiple scale analysis.

We collect terms of the same order in $\epsilon$ from Eq.~(\ref{apeqn3a}) after substituting Eqs.~(\ref{apeqn14})–(\ref{apeqn19}). At $\mathcal{O}(\epsilon)$, we obtain
\begin{equation}
\label{apeqn20}
L\phi_1 = \left(-a\frac{\partial^{2}}{\partial x^{2}} - K\frac{\partial^{4}}{\partial x^{4}} - \Gamma_c\right)\phi_{1} = 0.
\end{equation}
This corresponds to the linearized form of Eq.~(\ref{apeqn3a}) evaluated at the critical wave number $q_c$ and reaction rate $\Gamma_c$. At $\mathcal{O}(\epsilon^2)$, we get the following relation
\begin{equation}
\label{apeqn21}
L\phi_2 = -g\frac{\partial^2}{\partial x^2}\phi_1^2
-\lambda\frac{\partial^2}{\partial x^2}\left(\frac{\partial \phi_1}{\partial x}\right)^2
+\xi \frac{\partial}{\partial x}\left(\frac{\partial^2 \phi_1}{\partial x^2} \frac{\partial \phi_1}{\partial x}\right) + 2a\frac{\partial}{\partial x}\frac{\partial\phi_1}{\partial X} + 4K\frac{\partial^3}{\partial x^3}\frac{\partial\phi_1}{\partial X}.
\end{equation}
Substituting $\phi_1$ and $\phi_2$ from Eq.~(\ref{apeqn19}) into Eq.~(\ref{apeqn21}) and comparing the coefficients of $e^{i2q_cx}$, we obtain the following relation between $A(X,T)$ and $B(X,T)$: 
\begin{equation}
\label{apeqn22}
B(X,T) = \frac{4 A^2(X,T)}{9K} \left\{ -\frac{g}{q^2_c} + \left( \lambda - \frac{\xi}{2} \right) \right\}.
\end{equation}
Note that the last two terms on the right-hand side of Eq.~(\ref{apeqn21}) do not contribute, as they are proportional to $e^{iq_cx}$ or $e^{-iq_cx}$. Next, we compare the terms of $\mathcal{O}(\epsilon^3)$, which yields 
\begin{align}
\label{apeqn23}
\frac{\partial \phi_1}{\partial T} =& L\phi_3 + \left[\Gamma_c\phi_1 + u\frac{\partial^2\phi_1^3}{\partial x^2} + 2g \frac{\partial^2}{\partial x^2}(\phi_1\phi_2)
+ 2\lambda\frac{\partial^2}{\partial x^2}\left(\frac{\partial \phi_1}{\partial x}\frac{\partial \phi_2}{\partial x} 
 \right)
 -\xi\frac{\partial}{\partial x}\left(\frac{\partial^2 \phi_1}{\partial x^2}\frac{\partial \phi_2}{\partial x} 
+ \frac{\partial^2 \phi_2}{\partial x^2}\frac{\partial \phi_1}{\partial x}\right)\right. \nonumber \\
&
\left.-\left(a \frac{\partial^2 \phi_1}{\partial X^2} + 6K\frac{\partial^2}{\partial x^2}\frac{\partial^2 \phi_1}{\partial X^2}\right)\right] + \left\{2g\frac{\partial}{\partial x}\frac{\partial}{\partial X}\phi_1^2 - 4K\frac{\partial^3}{\partial x^3}\frac{\partial\phi_2}{\partial X} + 2\lambda\frac{\partial}{\partial x}\frac{\partial}{\partial X} \left(\frac{\partial\phi_1}{\partial x}\right)^2  \right. \nonumber \\ & \left. + 2\lambda \frac{\partial^2}{\partial x^2}\frac{\partial\phi_1}{\partial x} \frac{\partial\phi_1}{\partial X} - 2\xi \frac{\partial}{\partial x}\frac{\partial}{\partial x} \frac{\partial\phi_1}{\partial x} \frac{\partial\phi_1}{\partial X} - \xi \frac{\partial}{\partial x}\frac{\partial^2\phi_1}{\partial x^2}\frac{\partial\phi_1}{\partial X} - \xi \frac{\partial}{\partial X}\frac{\partial^2\phi_1}{\partial x^2}\frac{\partial\phi_1}{\partial x} \right\}.
\end{align}
Substituting $\phi_1$ and $\phi_2$ from Eq.~(\ref{apeqn19}) into Eq.~(\ref{apeqn23}) and comparing the coefficients of $e^{iq_cx}$, we obtain the following dynamical equation for $A(X,T)$:  
\begin{align}
\label{apeqn24}
\frac{\partial A}{\partial T} = \Gamma_c A 
+\frac{8q^4_c}{9K} \left[ \left\{ \frac{g}{q^2_c} + \frac{1}{2} \left( \lambda - \frac{\xi}{2} \right) \right\}^2 
- \frac{9}{4} \left\{  \left( \lambda - \frac{\xi}{2} \right)^2+\frac{3uK}{2q_c^2} \right\} \right] \lvert A \rvert^2 A 
+ 2a \frac{\partial^2 A}{\partial X^2}. 
\end{align}
This equation, known as the \textit{amplitude equation}, describes the spatiotemporal evolution of the dominant amplitude $A(X,T)$ of the order parameter $\phi(x,t)$ in the generalized AMB+ near the onset of instability. Note that in deriving Eq.~(\ref{apeqn24}), terms involving $B(X,T)$ have been replaced using Eq.~(\ref{apeqn22}). Also, the terms in \{....\} of Eq.~(\ref{apeqn23}) do not contribute, as they are not proportional to $e^{iq_cx}$. By comparing Eq.~(\ref{apeqn24}) with Eq.~(\ref{apeqn10}), the coefficients $\alpha_i$ can be identified as follows:
$$\alpha_1 = \Gamma_c, ~~\alpha_2 = \frac{8q^4_c}{9K} \left[ \left\{ \frac{g}{q^2_c} + \frac{1}{2} \left( \lambda - \frac{\xi}{2} \right) \right\}^2 
- \frac{9}{4} \left\{  \left( \lambda - \frac{\xi}{2} \right)^2+\frac{3uK}{2q_c^2} \right\} \right], ~~\alpha_3=2a.$$ Clearly, all $\alpha_i$'s are real quantity.

From Eq.~(\ref{apeqn24}), it is evident that the nature of transition at the onset of instability is supercritical when $\alpha_2 < 0$, i.e.,
\begin{align}
\label{apeqn28}
\left\{ \frac{g}{q^2_c} + \frac{1}{2} \left( \lambda - \frac{\xi}{2} \right) \right\}^2 <\frac{9}{4} \left\{  \left( \lambda - \frac{\xi}{2} \right)^2+\frac{3uK}{2q^2_c} \right\}.    
\end{align}
The transition becomes subcritical when $\alpha_2 > 0$, i.e., when the opposite inequality holds:
\begin{align}
\label{apeqn29}
\left\{ \frac{g}{q^2_c} + \frac{1}{2} \left( \lambda - \frac{\xi}{2} \right) \right\}^2>\frac{9}{4} \left\{  \left( \lambda - \frac{\xi}{2} \right)^2+\frac{3uK}{2q^2_c} \right\}.  
\end{align}

\begin{figure}
\centering
\includegraphics*[width=0.50\textwidth]{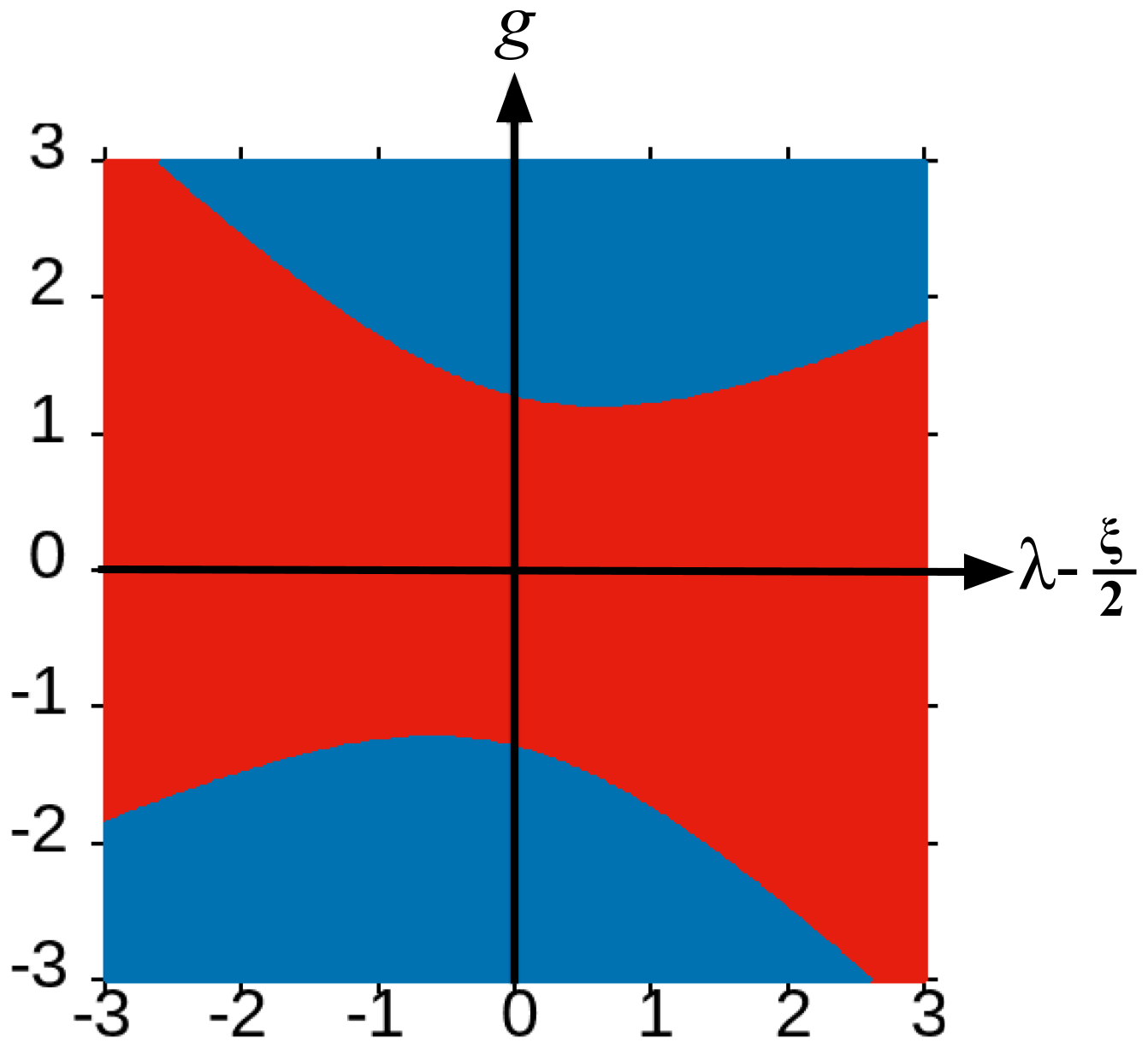}
\caption{\label{fig3a} Phase diagram marking the domains of supercritical-subcritical transition in the $(\lambda-\tfrac{\xi}{2})$~--~$g$ plane for $a=1$, $u=1$, and $K=1$. In the regions marked in red, supercritical transitions take place, while in the blue regions, subcritical transitions occur.}
\end{figure}

Figure~\ref{fig3a} shows the phase diagram indicating the domains of supercritical and subcritical transitions in the $(\lambda-\tfrac{\xi}{2})$~--~$g$ plane. The numerical values of the  parameters are provided in the figure caption.

\begin{figure}
\centering
\includegraphics*[width=0.90\textwidth]{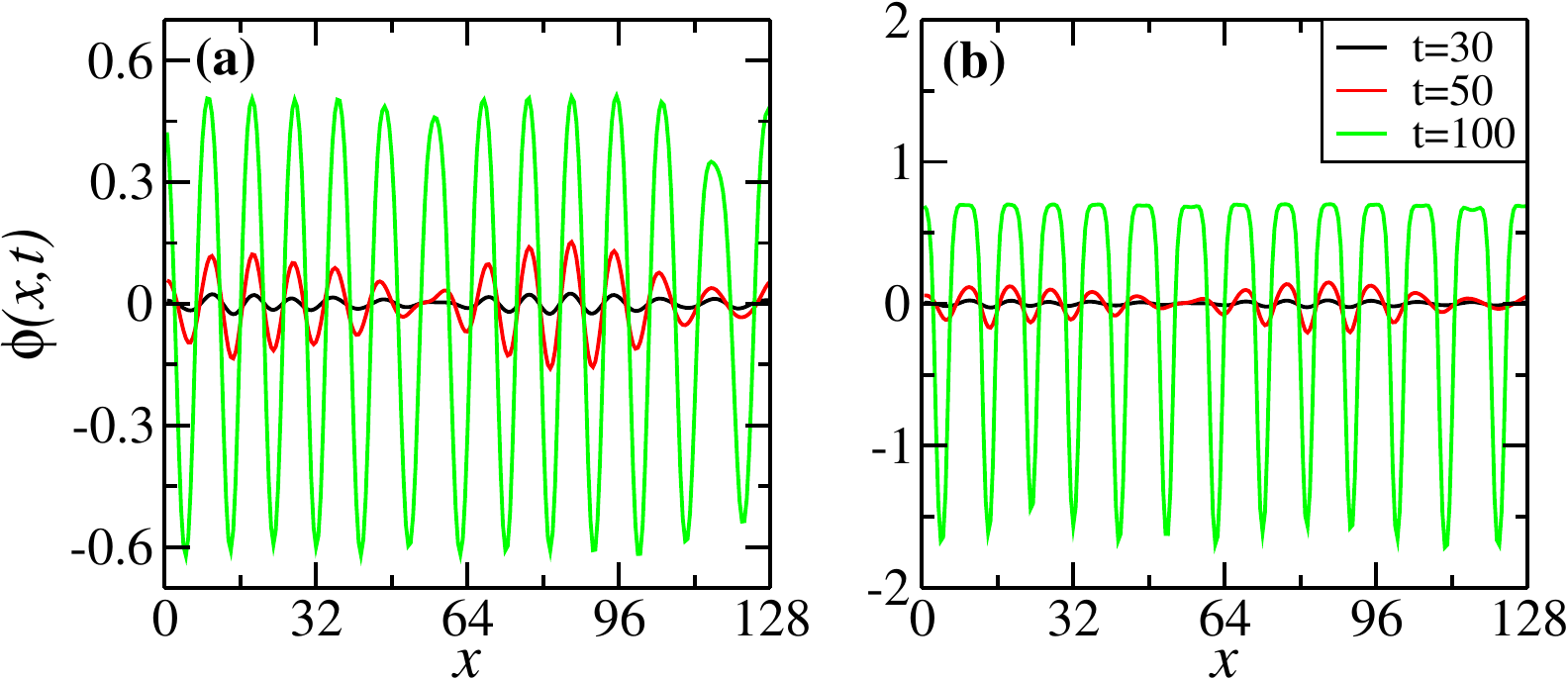}
\caption{\label{fig3} Time evolution of $\phi(x,t)$ in $d = 1$ for a system of size $L = 128$, obtained by solving Eq.~(\ref{apeqn3}) with parameters in (a) the supercritical regime: $a = 1$, $g = 0.5$, $\lambda = 0.2$, $\xi = 0.2$, $K = 1$, $u=1$, and $\Gamma = 0.15$; and (b) the subcritical regime: $a = 1$, $g = 2.0$, $\lambda = 0.2$, $\xi = 0.2$, $K = 1$, $u=1$, and $\Gamma = 0.15$. Lines of different colors indicate different times, as indicated.}
\end{figure}

Figure~\ref{fig3} shows $\phi(x,t)$ vs. $x$ in $d = 1$ at different times for (a) the supercritical regime and (b) the subcritical regime. The model parameters given in the figure caption are chosen such that the conditions for supercriticality [Eq.~(\ref{apeqn28})] and subcriticality [Eq.~(\ref{apeqn29})] are satisfied. In the supercritical regime, $\phi(x,t)$ exhibits harmonic profiles at the onset of instability, which grow in amplitude over time, as shown in Fig.~\ref{fig3}(a). In contrast, in the subcritical regime, $\phi(x,t)$ displays anharmonic profiles at the onset of instability, as shown in Fig.~\ref{fig3}(b).

\begin{figure}
\centering
\includegraphics*[width=0.90\textwidth]{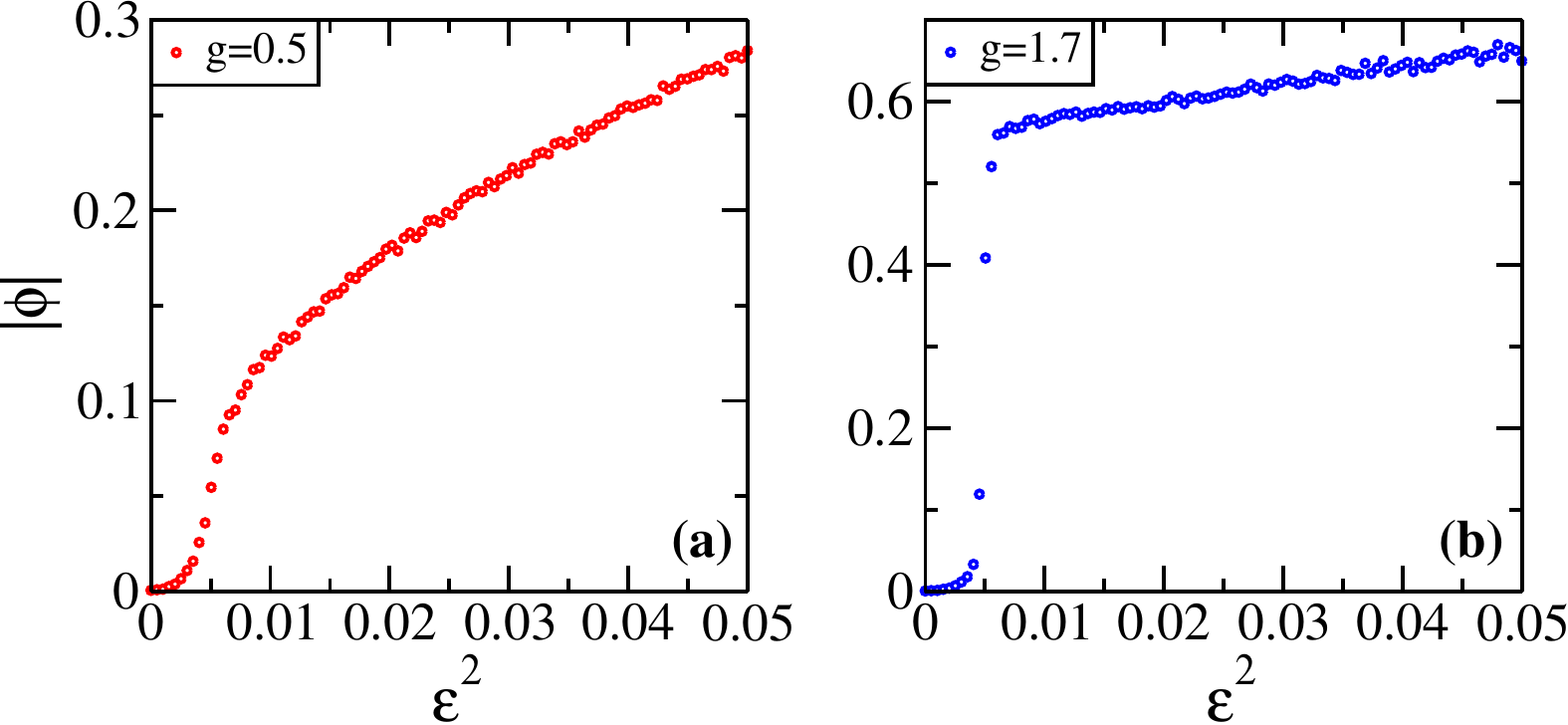}
\caption{\label{fig4} Plot of $|\phi|$ versus $\epsilon^2$ for parameters $a = 1$, $\lambda = 0.2$, $u=1$, $\xi = 0.2$, and $K = 1$ in (a) the supercritical regime with $g = 0.5$ and (b) the subcritical regime with $g = 1.7$. The critical value of $g$ is $g_c=1.275$ for the aforementioned parameters in the positive-$g$ domain.} 
\end{figure}

Next, we examine how the root-mean-squared amplitude of $\phi(x,t)$, denoted by $|\phi|$, varies as the system moves away from the onset of instability by changing $\epsilon$. It is given by
\begin{equation}
\label{apeqn26a}
|\phi|=\sqrt{\int_0^L\phi^2(x,t)dx}
\end{equation}
and represents a typical measure of the spatial fluctuation amplitude. Figures~\ref{fig4}(a) and \ref{fig4}(b) show the plots of $|\phi|$ versus $\epsilon^2$ in the supercritical and subcritical regimes, respectively. The numerical data are obtained by averaging over ten independent initial conditions, and the choice of parameters is given in the figure caption. Clearly, in the supercritical regime, $|\phi|$ increases continuously, whereas in the subcritical regime, the growth is discontinuous. Using Eq.~(\ref{apeqn24}), the homogeneous solution for the amplitude $A$ can be written as follows: 
\begin{equation}
\label{apeqn27}
|A| = \sqrt{\frac{\Gamma_c}{-\alpha_2}} = \sqrt{\frac{\Gamma_c}{ \frac{8q^4_c}{9} \left[ \left\{ \frac{g}{q^2_c} + \frac{1}{2} \left( \lambda - \frac{\xi}{2} \right) \right\}^2 
- \frac{9}{4} \left\{  \left( \lambda - \frac{\xi}{2} \right)^2+\frac{3uK}{2q^2_c} \right\} \right] }}.
\end{equation}
This expression suggests that the transition between the supercritical and subcritical regimes is controlled by $g$ and the activity parameters.

Next, we examine various limiting cases of the parameters $g$, $\lambda$, and $\xi$ to recover different models:\ \\
(a) For $g = 0$, $\lambda \ne 0$, and $\xi \ne 0$, Eq.~(\ref{apeqn3}) reduces to the AMB+ model with chemical reactions. In this case, $\alpha_2$ is always negative, indicating that the transition is supercritical.\ \\
(b) For $g = 0$, $\lambda \ne 0$, and $\xi = 0$, Eq.~(\ref{apeqn3}) reduces to the AMB model with chemical reactions, where $\alpha_2$ also remains negative, implying a supercritical transition.\ \\
(c) For $g = 0$, $\lambda = 0$, and $\xi = 0$, or for $g = 0$ and $\lambda = \xi/2$, Eq.~(\ref{apeqn3}) simplifies to the Cahn–Hilliard equation with chemical reactions, again yielding $\alpha_2 < 0$ and hence a supercritical transition. \ \\
Therefore, the parameter $g$ plays a crucial role in controlling the transition between the supercritical and subcritical regimes in the generalized AMB+. Finally,\ \\
(d) For $g \ne 0$, $\lambda = 0$, and $\xi = 0$, or for $g \ne 0$ and $\lambda = \xi/2$, Eq.~(\ref{apeqn3}) represents the Cahn–Hilliard equation for off-critical quenches with chemical reactions. In this case, $\alpha_2 = \left(\frac{8g^2}{9K} - 3 u q_c^2\right)$. Therefore, a transition between the supercritical and subcritical regimes occurs at $|g| = \sqrt{\frac{27ua}{16}}$. In general, the curve $\alpha_2(g)=0$, while keeping all other parameters fixed, marks the boundary between the subcritical and supercritical regimes in the parameter space.

\section{Analysis Of Amplitude Equations}\label{sec5}
To begin the analysis, we rewrite the amplitude equation describing roll patterns in the following form:
\begin{align}
\label{apeqn32}
 \frac{\partial A(X,T)}{\partial T} 
= \Gamma_c A(X,T) 
+ \alpha_2 \left(q_c, g, u, \left(\lambda - \tfrac{\xi}{2}\right)\right)\lvert A(X,T) \rvert^2  A(X,T) 
+ 2a\frac{\partial^2 A(X,T)}{\partial X^2}.
\end{align}
A supercritical transition requires $\alpha_2 < 0$; therefore, Eq.~(\ref{apeqn32}) can be rewritten in the supercritical regime as
\begin{align}
\label{apeqn32a}
 \frac{\partial A(X,T)}{\partial T} 
= \Gamma_c A(X,T) 
- \alpha \left(q_c, g, u, \left(\lambda - \tfrac{\xi}{2}\right)\right)\lvert A(X,T) \rvert^2  A(X,T) 
+ 2a\frac{\partial^2 A(X,T)}{\partial X^2},
\end{align}
where $\alpha=-\alpha_2>0$.

To study the stability of roll patterns with wavenumbers slightly shifted from the critical value $q_c$ in the steady state, we perturb the amplitude $A(X)$ as
\begin{align}
\label{apeqn33}
 A(X)= R_0 e^{iQX}.
\end{align}
where $R_0$ and $Q$ are constants. This is equivalent to saying that the steady-state order parameter is $\phi(x)\sim R_0 e^{i(q_cx+QX)}$. Substituting $A(X)$ from Eq.~(\ref{apeqn33}) into Eq.~(\ref{apeqn32a}) yields the relation
\begin{align}
\label{apeqn34}
 R^2_0=  \frac{\Gamma_c-2aQ^2}{\alpha}.
\end{align}
Since roll solutions can exist only when $R_0^2 > 0$, Eq.~(\ref{apeqn34}) implies that there is a band of permitted wavenumbers around $Q = 0$, corresponding to the critical wavenumber $q_c$.

To examine the stability of the roll patterns, we introduce small perturbations to the amplitude and phase of $A(X,T)$, such that $A(X,T) = \left(R_0 + r(X,T)\right)e^{i\left(QX + \Omega(X,T)\right)}$, where $|r|, |\Omega| \ll 1$. Substituting this into Eq.~(\ref{apeqn32a}) and linearizing in $r$ and $\Omega$, and then separating the real and imaginary parts, yields
\begin{align}
\label{apeqn35}
\frac{\partial r(X,T)}{\partial T}&= -2R^2_0 \alpha r -4QaR_0\frac{\partial \Omega}{\partial X}+2a \frac{\partial^2 r}{\partial^2 X},\ \\
\label{apeqn36}
R_0\frac{\partial \Omega(X,T)}{\partial T}&=4Qa \frac{\partial r}{\partial X}+ 2a R_0\frac{\partial^2 \Omega}{\partial^2 X}
\end{align}
As we are interested in long-wavelength effects, the spatial derivatives of the perturbation variables are small compared to the variables themselves. Therefore, we decompose $r$ and $\Omega$ into Fourier modes as $r = \tilde{r} e^{\tilde{\sigma}T + i l X}$ and $\Omega = \tilde{\Omega} e^{\tilde{\sigma}T + i l X}$. Substituting these ans\"atze into Eqs.~(\ref{apeqn35}) and (\ref{apeqn36}) yields the following relations:
\begin{align}
\label{apeqn35a}
&(2\alpha R^2_0+2al^2+\tilde{\sigma})\tilde{r} + 4iaR_0Ql\tilde{\Omega} = 0,\ \\
\label{apeqn36a}
& -4iaQl\tilde{r} + (2aR_0l^2+R_0\tilde{\sigma})\tilde{\Omega} = 0.
\end{align}
The growth rate $\tilde{\sigma}$ can be obtained by setting the determinant of the matrix
\begin{eqnarray}
\label{eqn37a}
M_q = \begin{bmatrix}
\rule{0pt}{2.5ex} 2\alpha R^2_0+2al^2+\tilde{\sigma} &  4iaR_0Ql \\ \rule{0pt}{2.5ex}
-4iaQl & 2aR_0l^2+R_0\tilde{\sigma}  \rule{0pt}{2.5ex}
\end{bmatrix}
\end{eqnarray}
to zero, which yields the following quadratic equation for $\tilde{\sigma}$:
\begin{align}
\label{apeqn37}
\tilde{\sigma}^2+ 2\tilde{\sigma}\left(\alpha R^2_0+2al^2\right)+4al^2\left(\alpha R^2_0+al^2-4aQ^2\right)=0
\end{align}
The two roots of Eq.~(\ref{apeqn37}) are given by
\begin{align}
\label{apeqn38a}
\tilde{\sigma}_\pm = -\left( \alpha R_0^2 + 2a l^2 \right) \pm \sqrt{\alpha^2 R_0^4 + 16 a^2 Q^2 l^2}.
\end{align}

The first root, $\tilde{\sigma}_-$, is given by
\begin{align}
\label{apeqn38}
\tilde{\sigma}_-=-2\alpha R^2_0+\mathcal{O}(l^2),
\end{align}
and the corresponding eigenvector is $(\tilde{r},\tilde{\Omega})=\left(1, \mathcal{O}(l^2)\right)$. Since $\tilde{\sigma}_-\approx-2\alpha R^2_0<0$, it indicates a relatively fast relaxation of the amplitude towards its equilibrium value. This implies that the roll state is stable with respect to amplitude perturbations.

The second root, $\tilde{\sigma}_+$, is given by
\begin{align}
\label{apeqn39}
\tilde{\sigma}_+ =-2al^2\left(1-\frac{4Q^2a}{\alpha R^2_0}\right)   + \mathcal{O}(l^4).
\end{align}
Clearly, $\tilde{\sigma}_+$ represents the genuine modulational mode, as it depends explicitly on $Q$ and $l$ at leading order, and the the corresponding eigenvector is $(\tilde{r},\tilde{\Omega})=\left(\mathcal{O}(l^2),1 \right)$. This mode describes the slow evolution of phase perturbations on a $\mathcal{O}(l^{-2})$ time scale. The phase instability, also known as the \textit{Eckhaus instability}, arises when $\tilde{\sigma}_+>0$, which is given by 
\begin{eqnarray}
\label{apeqn40aa}
Q^2 > \frac{R_0^2\alpha}{4a}.
\end{eqnarray}
This condition determines the stability boundary for roll patterns with respect to long-wavelength phase modulations. When the \textit{Eckhaus instability}~\cite{MP93,R06} criterion is satisfied, the roll pattern becomes unstable with respect to wavelength variations. If the roll wavelength is either too long or too short, the system responds by creating or annihilating rolls at defect points, thereby relaxing toward a more favorable and stable wavelength.

Using Eq.~(\ref{apeqn34}), we obtain the range of $Q^2$:
\begin{align}
\label{apeqn40a}
\frac{1}{3}\frac{\Gamma_c}{2a}<Q^2<\frac{\Gamma_c}{2a}.
\end{align}
for which the roll patterns become unstable solely due to phase modulations. Clearly, the boundaries are independent of the parameter $g$.

\section{Phase Diffusion Equation}\label{sec6}
As discussed earlier, the amplitude equation is invariant under a uniform phase shift, which implies that the dynamical equation for $\Omega$ will involve only spatial gradients. Furthermore, the reflection symmetry ($X \rightarrow -X$) of the underlying physical system allows $\frac{\partial^2 \Omega}{\partial X^2}$ to appear as the lowest-order term in the gradient expansion. Equation~(\ref{apeqn35}) shows that the damping term $-2\alpha R_0^2 r$ drives $r$ toward a value at which the RHS of the equation vanishes, and its steady-state value is determined by the relation
\begin{align}
\label{apeqn46}
\alpha R_0 r\simeq -2a Q\frac{\partial\Omega}{\partial X}.
\end{align}
This suggests that the amplitude dynamics is slaved to the phase dynamics, since the amplitude relaxes on a much faster time scale compared to the phase. Substituting Eq.~(\ref{apeqn46}) into Eq.~(\ref{apeqn36}), we obtain the dynamical equation of $\Omega(X,T)$ for generalized AMB+ in the presence of a chemical reaction:
\begin{align}
\label{apeqn47}
\frac{\partial \Omega}{\partial T}=\left(\frac{2a\alpha R^2_0-8a^2l^2}{\alpha R^2_0}\right)\frac{\partial^2 \Omega}{\partial X^2}=2a\left(\frac{\Gamma_c-6aQ^2}{\Gamma_c -2aQ^2}\right)\frac{\partial^2 \Omega}{\partial X^2}=D\frac{\partial^2 \Omega}{\partial X^2}.
\end{align}
Therefore, $\Omega(X,T)$ satisfies a diffusion equation, known as \textit{phase diffusion equation}, for patterns close to periodic stable states. The phase field $\Omega(X,T)$ becomes unstable to fluctuations when $D$ is negative, which occurs for $\tfrac{1}{3}\tfrac{\Gamma_c}{2a} < Q^2 < \tfrac{\Gamma_c}{2a}$. This corresponds exactly to the domain of the \textit{Eckhaus instability} given by Eq.~(\ref{apeqn40a}).

\section{Summary and Discussion}\label{sec7}
Let us conclude the results presented above with a brief summary and discussion. We have derived and analyzed the amplitude equation for roll patterns in the generalized AMB+ in the presence of a reversible chemical reaction with equal forward and backward rates $\Gamma$. The original AMB+, which describes phase separation in active binary mixtures, consists of a diffusion equation in which the current $\vec{J}(\vec{r}, t)$ has both passive and active components. The active current includes a rotation-free component of strength $\lambda$ and a rotational component of strength $\xi$, while the passive current can be derived from a free-energy functional. In the generalized AMB+, we introduce an additional quadratic term, $g\phi^2$, in the expression for the equilibrium part of the current. The chemical reaction removes the conservation constraint and introduces a preferred wavenumber that governs pattern formation below a critical reaction rate $\Gamma_c$. For $\Gamma > \Gamma_c$, the binary mixture remains homogeneous.

We first perform a linear stability analysis of the generalized AMB+, which allows us to identify the critical wavenumber $q_c$, the critical reaction rate $\Gamma_c$, and the band of wavenumbers that become unstable when $\Gamma < \Gamma_c$. However, this analysis does not reveal the nature of the transition at the critical point $\Gamma = \Gamma_c$, which is governed by the nonlinear terms in the AMB+. Although these nonlinear terms do not contribute to the linear stability analysis, they play a crucial role in saturating the growing modes at later times and in determining the nature of the transition near the onset. This behavior is captured through the derivation and analysis of the amplitude equation.

We argued for the form of the amplitude equation based on symmetry considerations and systematically derived it using multiscale analysis. By examining different limits of $g$, $\lambda$, and $\xi$, we obtained amplitude equations corresponding to several physical models. We found that for $g = 0$, the transition is always supercritical at the onset, whereas for $g \ne 0$, a transition between the supercritical and subcritical regimes can occur, depending on the other model parameters. Furthermore, we derived the condition for the \textit{Eckhaus instability} from the stability analysis of the amplitude equation, which defines a band of wavenumbers that become unstable solely due to phase fluctuations around $q_c$. Our analysis of the phase diffusion equation yields an identical band of unstable wavenumbers, independent of the parameter $g$. Thus, we conclude that $g$ influences the nature of the transition near the onset; however, the number of modes that become unstable when $\Gamma < \Gamma_c$ is independent of $g$.

\ \\
\noindent{\bf Acknowledgments:} SM acknowledge financial support from IISER Mohali through a Senior Research Fellowship.

\ \\
\noindent{\bf Conflict of interest:} The authors have no conflicts to disclose.

\end{document}